\documentclass{sig-alternate}
\newfont{\mycrnotice}{ptmr8t at 7pt}
\newfont{\myconfname}{ptmri8t at 7pt}
\let\crnotice\mycrnotice%
\let\confname\myconfname%

\permission{Permission to make digital or hard copies of all or part of this work for personal or classroom use is granted without fee provided that copies are not made or distributed for profit or commercial advantage and that copies bear this notice and the full citation on the first page. Copyrights for components of this work owned by others than ACM must be honored. Abstracting with credit is permitted. To copy otherwise, or republish, to post on servers or to redistribute to lists, requires prior specific permission and/or a fee. Request permissions from permissions@acm.org.}
\conferenceinfo{MM'14,}{November 3--7, 2014, Orlando, Florida, USA.}
\copyrightetc{Copyright 2014 ACM \the\acmcopyr}
\crdata{978-1-4503-3063-3/14/11\ ...\$15.00.\\
http://dx.doi.org/10.1145/2647868.2654933} 

\usepackage{cite}

\begin{document}

\title{Daily Stress Recognition from Mobile Phone Data, \\ 
Weather Conditions and Individual Traits}

\numberofauthors{6} 
%
\author{
%
%
\alignauthor
Andrey Bogomolov \\
\affaddr{University of Trento, \\SKIL Telecom Italia Lab}\\
\affaddr{Via Sommarive, 5}\\
\affaddr{I-38123 Povo - Trento, Italy}\\
\email{andrey.bogomolov@unitn.it}
\alignauthor
Bruno Lepri \\
\affaddr{Fondazione Bruno Kessler}\\
\affaddr{Via Sommarive, 18}\\
\affaddr{I-38123 Povo - Trento, Italy}\\
\email{lepri@fbk.eu}
\alignauthor
Michela Ferron \\
\affaddr{Fondazione Bruno Kessler}\\
\affaddr{via Sommarive, 18}\\
\affaddr{I-38123 Povo - Trento, Italy}\\
\email{ferron@fbk.eu}
\and
\alignauthor
Fabio Pianesi \\
\affaddr{Fondazione Bruno Kessler}\\
\affaddr{Via Sommarive, 18}\\
\affaddr{I-38123 Povo - Trento, Italy}\\
\email{pianesi@fbk.eu}
\alignauthor
Alex (Sandy) Pentland\\
\affaddr{MIT Media Lab}\\
\affaddr{20 Ames Street}\\
\affaddr{Cambridge, MA, USA}\\
\email{pentland@mit.edu}
}

\maketitle
\begin{abstract}
Research has proven that stress reduces quality of life and causes many
diseases. For this reason, several researchers devised stress detection systems
based on physiological parameters. However, these systems require that obtrusive
sensors are continuously carried by the user. In our paper, we propose an
alternative approach providing evidence that daily stress can be reliably
recognized based on behavioral metrics, derived from the user's mobile phone
activity and from additional indicators, such as the weather conditions 
(data pertaining to transitory properties of the environment) and the personality
traits (data concerning permanent dispositions of individuals). 
Our multifactorial statistical model, which is person-independent, obtains the
accuracy score of 72.28\% for a 2-class daily stress recognition problem. The model is efficient to implement
for most of multimedia applications due to highly reduced low-dimensional
feature space (32d).
Moreover, we identify and discuss the indicators which have strong predictive power. 

\end{abstract}




\section{Introduction}
Nowadays, the number of mobile phones in use worldwide is about 5 billion, with millions of new subscribers everyday \footnote{http://www.ericsson.com/ericsson-mobility-report}. Mobile phones allow for unobtrusive and cost-efficient access to
huge streams of previously inaccessible data related to daily social behavior
\cite{LaneEtAl2010}.
These devices are able to sense a wealth of behavioral data such as (i)
location, (ii) other devices in physical proximity through Bluetooth scanning,
(iii) communication data, including both metadata (logs of who, when, and
duration) of phone calls and text messages (sms), etc. Correspondingly, the availability is continuously growing of huge streams of \emph{personal data} related to activities, routines and social interactions \cite{LaneEtAl2010, DBLP:conf/mum/DongLP11} which represent a novel opportunity to address fundamental problems of our societies in different fields, such as mobility and urban planning \cite{Gonzalez2008}, finance \cite{Singh2013}, healthy living and subjective well-being \cite{MadanEtAL2012, LiKamWa:2013:MBM:2462456.2464449}.

In this work, we focus on one of the most widespread and
debilitating problem of our subjective well-being and our society: stress.
Stress is a well-known condition in modern life and research has shown that the
amount of cumulative stress plays a role in a broad range of physical,
psychological and behavioural conditions, such as anxiety, low self-esteem,
depression, social isolation, cognitive impairments, sleep and
immunological disorders, neurodegenerative diseases and other medical conditions
\cite{CohenS:1997}, 
while also significantly contributing to healthcare costs.
Hence, measuring stress in daily life situations has become an important
challenge \cite{5779068}. Today, the availability of huge and diverse streams of
pervasive data produced by and about people allows for automatic, unobtrusive,
and fast recognition of daily stress levels. An early prediction of stress symptoms can
indeed help to prevent situations that are risky for human life\cite{1438384}.

Several studies have produced interesting results that support the feasibility of
detecting stress levels through physiological sensors (see 
\cite{1438384}, \cite{1213626}). 
However, the use of physiological sensors is limited by several
shortcomings. Stress detection systems based on physiological measurement such as
heart-rate variability or skin conductance are intrusive and need to be easily
wearable to be exploited in natural settings; the data they produce can be confounded by
daily life activities such as speaking or drinking; they exhibit important between-person
differences \cite{5779068}.

Recently, social psychologist Miller wrote ``The Smartphone Psychology
Manifesto" in which he argued that the smartphones should be seriously
considered as new research tools for psychology. In his opinion, these tools
could revolutionize all fields of psychology and other behavioral sciences
making these disciplines more powerful, sophisticated, and grounded in
real-world behavior \cite{Miller01052012} and \cite{LathiaRachuri2013}. Indeed, several works have started to
use smartphone activity data in order to detect and predict personality traits
\cite{de2011towards, staiano2012friends, Chittaranjan:2013:MLS:2444516.2444522, Montjoye2013}, mood
states \cite{LiKamWa:2013:MBM:2462456.2464449}, and daily happiness
\cite{Muaremi_UbiHealth2012}. Stopczynski et al. \cite{Stopczynski2014} described the Copenhagen Networks Study, a large-scale study designed to measure human interactions spanning multiple years.

Smartphones data can be used to detect stress levels as well. Indeed, stress
levels are associated with the type of activities people engage in, including
those executed at/through their smartphone (for instance, a high number of phone
calls and/or e-mails from many different people could be associated with higher
stress levels). Weather conditions -- an environmental transitory property -- in
turn, have been argued \cite{howarth1984multidimensional},
\cite{sanders1982relationships} to be  often associated with stress, acting
either directly (as stressors) or indirectly (by affecting individual sensitivity to
stressors). Finally, the impact of all these transitory factors  -- 
(smartphone) activities and weather conditions -- on stress induction can be expected to be
modulated by personal characteristics and differences  \cite{Suls01021998},
\cite{VollrathTorgersen2000}. For example, a neurotic person could react with
higher levels of stress to a high number of interactions (call, sms or proximity
interactions) than an emotionally stable person; an extrovert or agreeable
person, in turn, might well find him/herself at ease with a high number of
interactions.

In this paper, we approach the automatic recognition of daily stress as a
2-class classification problem (non-stressed vs stressed) based on information
concerning different types of data: a) people activities, as detected through
their smartphones; b) weather conditions; c) personality traits. The information
about people activities is represented by features extracted from call and sms
logs and from Bluetooth hits, able to capture (i) the amount of calls, of sms and of
proximity interactions; (ii) the diversity of calls, of sms, and of proximity interactions; and (iii) regularity in
user behaviors.
In addition, we use weather conditions (environmental and transitory factors)
along with personality traits (internal and stable factors); the latter are
mediating factors that can modulate people responses to stressors (e.g.,
weather, daily activity).
This multifactorial approach will be compared to approaches based only on a
family of features (personality, weather conditions, mobile phone features) or
 simpler combinations of families of features (personality and weather
conditions; personality and mobile phone features; weather conditions and mobile
phone features).

Classification experiments are performed using a variety of approaches and the
best solution for our classification problem was found using an ensemble of tree
classifiers based on a Random Forest algorithm. Our multifactorial approach
obtains an accuracy score of 72.28\% for a 2-class daily stress recognition
problem, providing evidence that individual daily stress can be reliably
predicted from the combination of smartphone usage data, weather conditions and individual dispositions
(personality traits). Interestingly, if one of these information sources is
dropped, the recognition performances decrease drastically.

In sum, the main contributions of this paper are as follows:
\begin{enumerate}
\item We propose a multi-factorial data-driven approach to the prediction of individual daily stress;
\item We validate our approach with a seven-months dataset collected from 111 subjects;
\item We provide a comprehensive analysis of the predictive power of the proposed approach and a comparison with approaches based only on single families of features (personality, weather conditions, mobile phone features) or pairwise combinations thereof  (personality and weather conditions, personality and mobile phone features, weather conditions and mobile phone features).
\end{enumerate}

\section{Related Work}
A large body of research on stress detection focused on physiological
measurements to infer stress levels (see \cite{1438384},
\cite{5704068}, \cite{5779068}).
Heart-rate variability, galvanic skin response, respiration, muscle activity and
temperature are among the most relevant features. However, despite providing
reliable insights on stress levels, this approach has major limitations because
it comprises wearable sensors that need to be carried at all times to allow for
continuous monitoring.

Among the different changes in physiological parameters that happen during
stressful situations, variation in speech production has inspired a number of
studies using acoustic sensing on smartphones. Research on stress detection
based on voice analysis considered different speech characteristics such as
pitch, glottal pulse, spectral slope and phonetic variations. For example, Lu
and colleagues \cite{Lu:2012:SDS:2370216.2370270} proposed StressSense, an
Android application for stress detection from human voice in real-life
conversation, and they achieved 81\% and 76\% accuracy for indoor and outdoor
environments.

However, these methods depend on sound quality, which is not granted in
natural settings (e.g., crowded public places, noisy outdoor), and the
correlation between speech and emotion is subjected to large individual
differences \cite{scherer2002acoustic}. Hence, our performance of 72.28\% is a good and reliable
alternative to stress detection.
Other studies focused on the video analysis of behavioural correlates of
psychological stress \cite{giakoumis2012real}. These systems, despite providing
an unobtrusive method for stress monitoring, cannot be employed in a large
variety of real world and mobile environments and pose privacy concerns related
to the recording of people's behaviour.

A promising approach that can overcome the major shortcomings of stress
detection based on physiological measures and on audio/video analysis is
activity recognition from smartphone usage patterns. Studies in this field have
been mainly focused on the understanding of relational dynamics of individuals
\cite{eagle2009inferring}. 
Recently studies have started to investigate how smartphone usage habits can
provide insights into users' affective state
\cite{LiKamWa:2013:MBM:2462456.2464449} and stress levels \cite{bauer2012can}.
LiKamWa and colleagues \cite{LiKamWa:2013:MBM:2462456.2464449} proposed
MoodScope, a mobile software system that recognizes the users' mood, but not
stress states, from smartphone usage analysis. They collected usage data and
self-reported mood in a two months longitudinal study and used them to train
mood models. Smartphone usage data consisted in phone calls, SMSes, e-mail
messages, application use, web browsing histories and location changes, while
self-reported mood was collected from users' input at least four times a day.
MoodScope reached a 66\% accuracy of participants' daily-average mood, with
phone calls and categorized applications as the most useful features for mood
discrimination.

Bauer and Lukowicz \cite{bauer2012can} focused on mid-term stress detection,
monitoring 7 students during a two week exam session followed by two weeks of
non-stressful period. The recorded data were related to participants' mobility
patterns and social interactions, and included users' location, Bluetooth
proximity, phone calls and SMSes. These features allowed to detect an average
behaviour modification of 53\% for each user during the exam session. A
limitation of this study is the small number of subjects. Our multifactorial
approach outperforms the approach proposed by \cite{bauer2012can} although a
direct comparison may be not adequate given the different focus: our approach
tend to daily classify people as "not stressed" or "stressed", while Bauer and
Lukowicz try to detect stressful situations.

In 2013, Sano and Picard \cite{SanoPicard2013} reported an accuracy performance
in stress recognition of 75\% using a combination of features obtained from
mobile phones and wearable sensors. However, the limited number of subjects used
in their experiments (18) and the limited number of days (5) make preliminary
the results of this study.

\section{Data Collection}
From November 12, 2010 to May 21, 2011, we collected a dataset capturing 
the lives of
117 subjects living in a married graduate student residency of a major US
university. Our sample of subjects has a large variety in terms of provenance and cultural background: we have subjects from 16 countries such as USA, China, Israel, India, Iran, Russia, etc.  During this period, each participant was equipped with an
Android-based cellular phone incorporating a sensing software explicitly
designed for collecting mobile data. Such software runs in a passive manner and
does not interfere with the every day usage of the phone.
The data collected consisted of: (a) call logs, (b) sms logs, (c) proximity data, obtained by scanning near-by phones and other Bluetooth devices
every five minutes, and (d) data from surveys administered to participants,
which provided self-reported information about personality traits (``Big Five") and %
self reported information about daily stress.

Proximity interaction data were derived from Bluetooth hits in a
similar way as in previous reality mining studies
\cite{DBLP:journals/puc/EagleP06}.  Bluetooth scans
were performed every 5 minutes in order to keep  the battery from draining while
achieving a high enough temporal resolution. The Bluetooth log of a given smartphone were then used to extract the  list of the other 
participants' phones which were in proximity.

In total, the dataset consisted of 33497 phone calls, 22587 SMS, and 1460939
Bluetooth hits.

\subsection{Stress data}
At the evening, the participants were also asked to fill daily surveys about their
daily self-perceived stress level. The stress information was reported by the
participants filling a seven items scale with 1 = ``not stressed", 4 = ``neutral"
and 7 = ``extremely stressed". In our experiments we used the data only for the subjects (111 subjects) who had provided at least 2 weeks of consecutive data.

\begin{figure}[ht] \centering{
\includegraphics[scale=0.5]{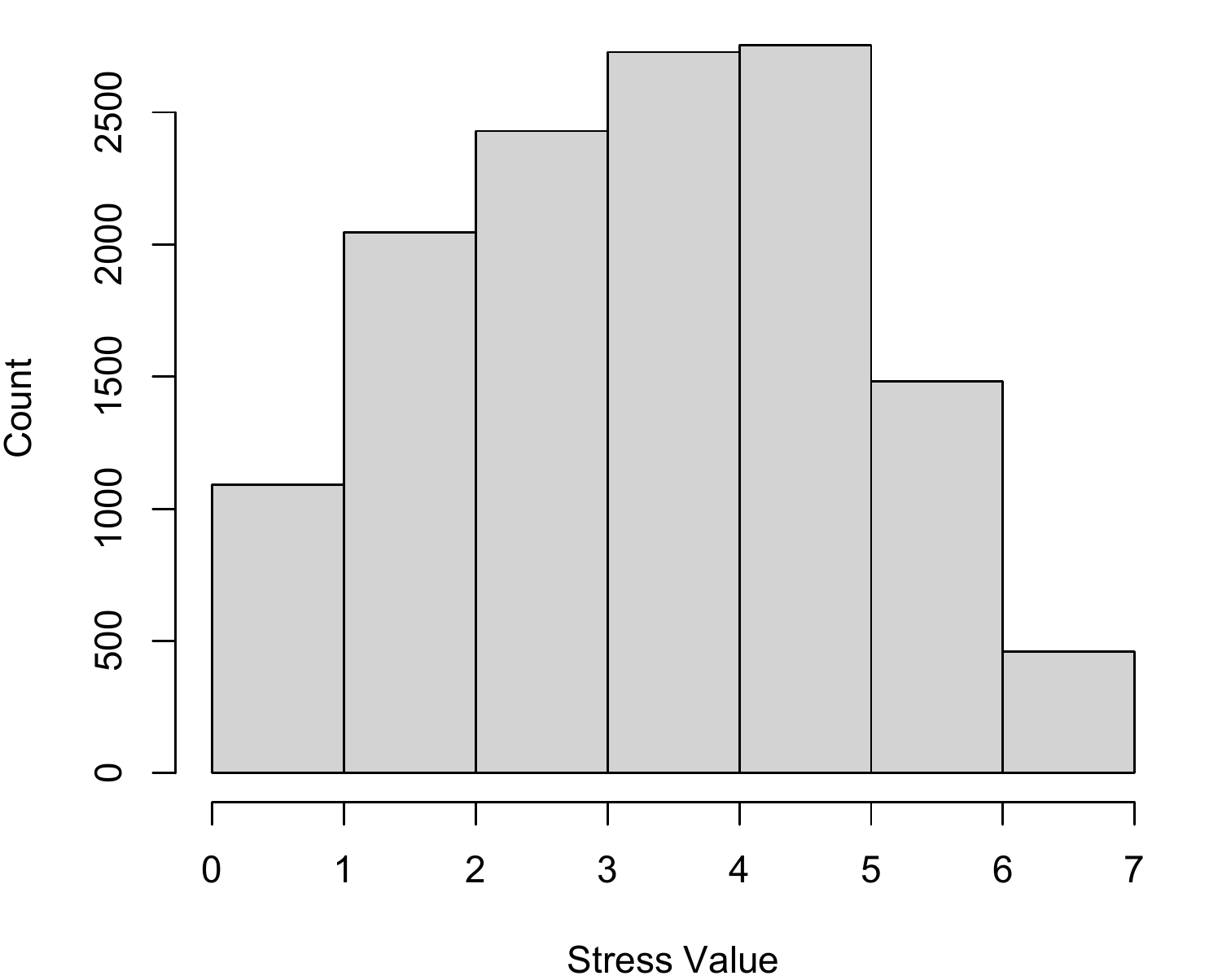}}
\caption{Recorded Stress Scores Density}
\label{fig:stressDescriptiveStatistics}
\end{figure}

\begin{figure}[ht] \centering{
\includegraphics[scale=0.55]{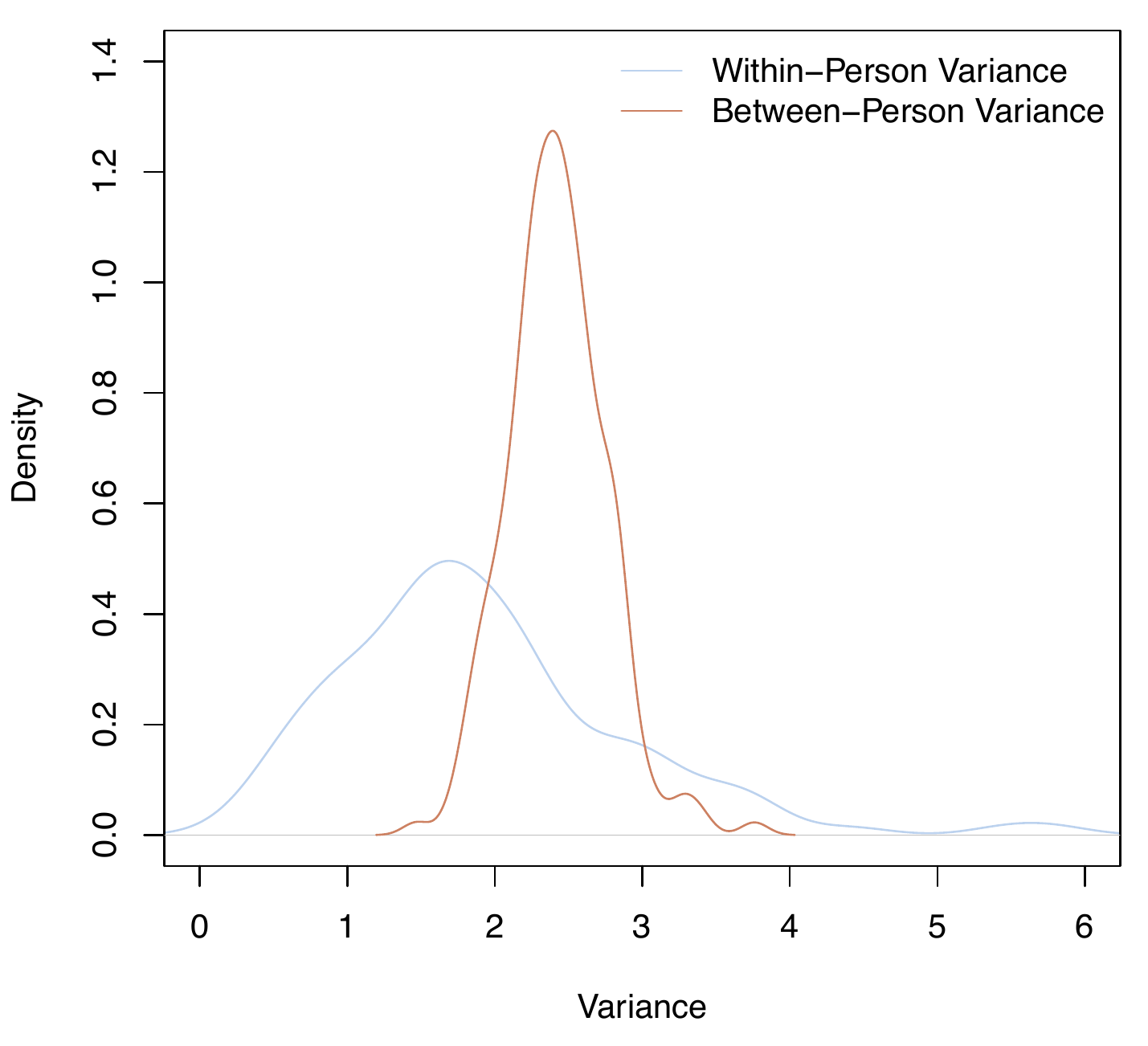}}
\caption{Within- and between-subject variance}
\label{fig:withinAndBetweenPersonVariance}
\end{figure}

The distribution of daily stress is visualized in
Fig.~\ref{fig:stressDescriptiveStatistics}. We see that it has a small negative
skew -- the density is moved to the higher region of stress score.
The distribution has negative excess kurtosis, which in our case means that the sample reported a specific
daily stress score more often than the neutral. 
Fig.~\ref{fig:withinAndBetweenPersonVariance} shows that within-person daily
stress variance is more spread than between-person, but the density of between-person variance is higher.

\begin{table*}
\caption{Selected Features Ranked by Mean Decrease in Accuracy} 
\label{tab:top30features}
\centering
\scalebox{0.8}{
\begin{tabular}{rl|rrrr}
  \hline
Rank & Feature & 0 & 1 & Mean Decrease in Accuracy & Mean Decrease in Gini Index \\
  \hline
1 & personality.Conscientiousness & 13.65 & 18.04 & 23.35 & 159.96 \\ 
  2 & personality.Agreeableness & 14.22 & 19.73 & 22.92 & 167.30 \\ 
  3 & personality.Neuroticism & 15.96 & 21.04 & 22.56 & 183.87 \\ 
  4 & personality.Openness & 14.20 & 14.18 & 21.38 & 139.23 \\ 
  5 & personality.Extraversion & 15.75 & 15.02 & 21.07 & 158.51 \\ 
  6 & weather.MeanTemperature & 14.50 & 6.34 & 17.44 & 322.27 \\ 
  7 & sms.RepliedEvents.Latency.Median & 8.83 & 13.85 & 15.63 & 48.74 \\ 
  8 & weather.Humidity & 15.33 & 2.10 & 15.45 & 298.13 \\ 
  9 & sms.AllEventsPerDay & 8.61 & 0.56 & 10.50 & 42.91 \\ 
  10 & bluetooth.Q95TimeForWhichIdSeen & 4.99 & 6.05 & 9.94 & 32.47 \\ 
  11 & bluetooth.MaxTimeForWhichIdSeen & 6.24 & 7.23 & 9.47 & 32.12 \\ 
  12 & sms.IncomingAndOutgoingPerDay & 7.45 & 1.26 & 9.38 & 41.59 \\ 
  13 & weather.Visibility & 9.94 & 1.26 & 9.22 & 251.27 \\ 
  14 & weather.WindSpeed & 8.77 & 1.30 & 8.67 & 282.10 \\ 
  15 & bluetooth.Q90TimeForWhichIdSeen & 4.24 & 6.75 & 8.64 & 28.41 \\ 
  16 & bluetooth.TotalEntropyShannon & 5.04 & 3.51 & 8.56 & 31.37 \\ 
  17 & call.EntropyMillerMadowOutgoingTotal & 4.25 & 4.10 & 8.54 & 27.49 \\ 
  18 & call.EntropyShannonOutgoingAndIncomingTotal & 4.23 & 4.86 & 8.53 & 26.28
  \\
  19 & bluetooth.TotalEntropyMillerMadow & 5.06 & 4.22 & 8.50 & 32.09 \\ 
  20 & bluetooth.IdsMoreThan09TimeSlotsSeen & 6.11 & 5.85 & 8.43 & 27.88 \\ 
  21 & bluetooth.IdsMoreThan04TimeSlotsSeen & 6.34 & 4.59 & 8.04 & 24.64 \\ 
  22 & call.EntropyShannonMissedOutgoingTotal & 3.13 & 4.92 & 7.85 & 24.34 \\ 
  23 & bluetooth.IdsMoreThan19TimeSlotsSeen & 2.97 & 5.16 & 7.78 & 20.87 \\ 
  24 & call.EntropyShannonOutgoingTotal & 3.10 & 6.45 & 7.78 & 24.79 \\ 
  25 & bluetooth.Q75TimeForWhichIdSeen & 5.16 & 4.70 & 7.76 & 22.07 \\ 
  26 & call.EntropyMillerMadowMissedOutgoingTotal & 4.09 & 5.45 & 7.55 & 24.64
  \\
  27 & call.EntropyMillerMadowOutgoingAndIncomingTotal & 3.87 & 6.29 & 7.51 &
  28.63 \\
  28 & sms.OutgoingAndIncomingTotalEntropyMillerMadow & 4.68 & 3.84 & 7.19 &
  17.63 \\
  29 & sms.OutgoingTotalEntropyMillerMadow & 5.22 & 1.49 & 7.19 & 18.88 \\ 
  30 & bluetooth.Q50TimeForWhichIdSeen & 1.53 & 7.29 & 7.08 & 18.91 \\ 
  31 & bluetooth.Q68TimeForWhichIdSeen & 2.36 & 5.96 & 6.68 & 19.05 \\ 
  32 & sms.OutgoingTotalEntropyShannon & 2.53 & 2.77 & 5.13 & 17.59 \\ 
   \hline
\end{tabular}
}
\end{table*}

\subsection{Personality Data}
Several studies in social psychology investigated the relationships between 
personality traits and psychological stress. Personalities that tend to be more
negative are usually associated with greater distress, while outgoing and
positive personalities generally experience less distress \cite{Suls01021998},
\cite{VollrathTorgersen2000}. The majority of the studies that have examined the
relationship between personality and distress focused on the Big Five traits
\cite{John1999}, a  personality model owing its name to the  five traits it takes as a constitutive
of people's personality:
Extraversion, Neuroticism, Agreeableness, Conscientiousness, and
Openness to experience. Researchers showed significant associations  between psychological stress, on the one hand, and Neuroticism,
Extraversion and Conscientiousness, on the other. 
Duggan et al. \cite{Duggan1995139} found that individuals  high in Neuroticism
may be more vulnerable to experiencing distress as they respond more negatively
to daily stressors and report more daily stressful events and higher levels of
daily stress. When people with high scores in Neuroticism encounter stressful
events, they tend to experience them as more aversive than those low in this
trait \cite{JOPY:JOPY355}, \cite{Gunthert99}. Finally, in a study with
university students, Volrath and Torgensen \cite{VollrathTorgersen2000} showed
that students with more adaptive personalities such as Extraversion and
Conscientiousness are more positive and sociable and hence less affected by
daily stress.

In our study, Big Five personality traits were measured by asking subjects to answer the online version of the 44 questions Big Five questionnaire
developed by John et al. \cite{John1999}, by means of 5-point likert scales.  The scores on the five traits were the average over
the raw scores (inverted when needed) of the items pertaining 
to each trait.

\subsection{Weather data}
 The question about the relationship between mood, health and weather has been extensively debated 
\cite{hardt1999no},\cite{sanders1982relationships}. Studies in environmental psychology investigated the role of weather as a stressor and showed  significant effects of temperature, hours of sunshine and humidity on mood 
\cite{howarth1984multidimensional}, \cite{sanders1982relationships}. 
A large-scale study by Faust and colleagues \cite{faust1974influence} on 16,000 students in Switzerland showed an association between weather and sleep disorders, depressed mood and irritability. 
More recently, Denissen et al. \cite{3f049b980c4b47e8b3365140181f5062}
investigated the effects of six daily weather variables (temperature, wind
power, sunlight, precipitation, air pressure, photoperiod) on three mood
variables (positive affect, negative affect, and tiredness). 
Their results revealed
main effects of temperature, wind power, and sunlight on negative affect, while sunlight had also a main effect on tiredness and mediated the effects of
precipitation and air pressure on tiredness.

In our experiments, we used the following weather variables: (i) mean temperature, (ii)
pressure, (iii) total precipitation, (iv) humidity, (v) visibility and (vi) wind speed (measured in m/s) metrics.
The source weather data were collected from Wolfram Alpha \footnote{http://www.wolframalpha.com}. All weather metrics are computed on a daily scale for the same day that is under investigation and the source data are extracted from the Boston area weather stations (e.g. KBOS) located on the same relative elevation as the campus where the data collection was performed.

\section{Feature Extraction}
\label{sec:featureExtraction}
Based on previous works that characterize
social interactions by means of mobile phone data and use social interactions data to predict people's behaviors, states \cite{BogomolovSocialCom2013, LiKamWa:2013:MBM:2462456.2464449}, and traits \cite{de2011towards, Chittaranjan:2013:MLS:2444516.2444522, Montjoye2013}, 
we derived
the 25 call and sms basic features reported in
Table~\ref{tab:ListOfExtractedFeatures} and the 9 proximity basic features reported in Table~\ref{tab:ListOfBTFeatures}.

For each basic feature, we calculated second order features, such as mean, median, min, max,
99\%, 95\% quantiles, quantiles corresponding to 0.5, 1, 1.5 and 2 standard
deviations (applying Chebyshev's inequality), variance and
standard deviation functions. Moreover, for each basic feature  we calculated the same
functions as above for 2 and 3 days backward-moving window to account for the possibility that past events influenced the current stress state.

In the following subsections we will describe more in detail the 25 call and sms basic features and the 9 proximity basic features.

\subsection{Call and Sms Features}
The features reported in Table~\ref{tab:ListOfExtractedFeatures} fall
under four broad categories: (i) general phone usage, (ii) active behaviors, (iii) regularity, and (iv) diversity.
\begin{table}[ht]
	\caption{List of Basic Features}
	\label{tab:ListOfExtractedFeatures}
	\centering
	\scalebox{0.8}{
		\begin{tabular}{l}
			\hline
			General Phone Usage\\
			\hline
1.	Total Number of Calls (Outgoing+Incoming)\\
2.	Total Number of Incoming Calls\\
3.	Total Number of Outgoing Calls\\
4.      Total Number of Missed Calls\\
5.      Number of SMS received\\
6.      Number of SMS sent\\
\hline
Diversity\\
\hline
7.      Number of Unique Contacts Called\\
8.      Number of Unique Contacts who Called\\
9.      Number of Unique Contacts Communicated with (Incoming+Outgoing)\\
10.      Number of Unique Contacts Associated with Missed Calls\\
11.	Entropy of Call Contacts\\
12.	Call Contacts to Interactions Ratio\\
13.     Number of Unique Contacts SMS received from\\
14.     Number of Unique Contacts SMS sent to\\
15.     Entropy of SMS Contacts\\
16.     Sms Contacts to Interactions Ratio\\
\hline
Active Behaviors\\
\hline
17.	Percent Call During the Night\\
18.	Percent Call Initiated\\
19.     Sms response rate\\
20.     Sms response latency\\
21.     Percent SMS Initiated\\
\hline
Regularity\\
\hline
22.	Average Inter-event Time for Calls (time elapsed between two events)\\
23.	Average Inter-event Time for SMS (time elapsed between two events)\\
24.	Variance Inter-event Time for Calls (time elapsed between two events)\\
25.	Variance Inter-event Time for SMS (time elapsed between two events)\\
			\hline
		\end{tabular}}
\end{table}

Features for \emph{general phone usage} consist of: the total number of outgoing, incoming and missed calls and the total
number of sent/received sms. Moreover, we also computed the following ratios:  outgoing to incoming
calls, missed to (outgoing + incoming) calls,  sent to received sms.

Then, we captured the \emph{active behaviors} of an individual computing the following features: (i) percentage of calls done during the night, (ii) percentage of initiated calls during the night, (iii) the sms response rate, (iv) the sms response latency, and (v) the percentage of initiated sms. In particular, we consider a text from a user (A) to be a response to a text received from another user (B) if it is sent within an hour after user A received the last text from user B. The response rate is the percentage of texts people respond to. The latency is the median time it takes people to answer a text. Note that by definition, latency will be less or equal to one hour.

\emph{Diversity} and \emph{regularity} have been shown to be important for the characterization of different facets of human behavior. In particular, entropy, used as a measure of diversity, has been  successfully applied to  predict  mobility \cite{Song2010}, spending patterns \cite{Krumme2013, Singh2013}, online behavior \cite{SinatraSzell2014} and  personality traits \cite{Montjoye2013}. Concerning \emph{regularity} features, we measured the time elapsed between calls, the
time elapsed between sms exchanges and the time elapsed between call and sms.
More precisely, we consider both the average and the variance of the inter-event
time of one's call, sms and sum thereof (call+sms). Noticeably, in fact,  even when two users
have the same inter-event time for both call and sms, that quantity can be 
different for their sum.

\emph{Diversity} measures how evenly an individual's time is distributed among others.
In our case, the diversity of user behavior is addressed by means of three kinds
of features:
(i) entropy of contacts, (ii) unique contacts to interactions ratio, (iii)
number of unique contacts,  all computed  both on calls and on
sms. In particular, the entropy of an individual is the ratio between his/her total
number of contacts and the relative frequency at which he/she interacts with
them. The more one interacts equally often with a large number of contacts, the
higher the entropy will be. For entropy calculation, we applied
{\em Miller-Madow correction}\cite{Miller:1955}, which
is explained in Equation \ref{eq:mmEntropy}.
\noindent
\begin{equation}
\label{eq:mmEntropy}
\hat{H}_{MM}(\theta) \equiv -\sum\limits_{i=1}^p{\theta_{ML,i}}\log{\theta_{ML,i}} + \frac{\hat{m}-1}{2N},
\end{equation}
where $\hat{m}$ is a number of bins with nonzero $\theta$-probability. The
likelihood function is given as the product of probability density functions 
$P(\theta)=f(x_1;\theta)f(x_2;\theta)\cdots f(x_n;\theta)$ for a random sample
$X_1,\cdots,X_n$  . $\theta_{ML}$ is the maximum likelihood estimate of
$\theta$, which maximizes $P(\theta)$.
Miller-Madow correction was applied, dealing with the data quality problems, to
get bias-corrected empirical entropy estimate.

\subsection{Proximity Features}

Starting from the Bluetooth hits collected, we filtered out all the cases with $RSSI<0$.   
From the filtered Bluetooth proximity data
we extracted the following basic Bluetooth proximity features
(Table~\ref{tab:ListOfBTFeatures}).
\begin{table}[ht]
	\caption{List of Basic Bluetooth Proximity Features}
	\label{tab:ListOfBTFeatures}
	\centering
	\scalebox{0.8}{
		\begin{tabular}{l}
			\hline
			General Bluetooth Proximity\\
			\hline
1. Number of Bluetooth IDs\\
2. Times most common Bluetooth ID is seen\\
3. Bluetooth IDs accounting for n\% of IDs seen\\
4. Bluetooth IDs seen for more than k time slots\\
5. Time interval for which a Bluetooth ID is seen\\
6. Entropy of Bluetooth contacts\\
\hline
Diversity\\
\hline
7. Contacts to interactions ratio\\
\hline
Regularity\\
\hline
8. Average Bluetooth interactions inter-event time\\
(time elapsed between two events)\\
9. Variance of the Bluetooth interactions inter-event time \\
(time elapsed between two events)\\
			\hline
		\end{tabular}}
\end{table}
In this case, the extracted features fall under three broad categories: (i)
general proximity information, (ii) diversity, and (iii) regularity.
As for call and sms, we applied Miller-Madow correction for entropy calculation.

\section{Methodology}
We formulated the automatic recognition of daily stress as a binary classification
problem (``not stressed'' vs ``stressed''), with labels 
 $0$ for ``not stressed" and label $1$ for ``stressed". The two classes included all the cases with scores $<=4$ and scores $>4$, respectively. The sizes of the resulting two classes
are 36.16\% for "stressed" and 63.84\% for "not stressed". The inclusion of the cases with stress=4 in the $0$ class meant to provide a more clearcut distinction between the ``stressed" and the ``non-stressed" cases.

The data set was then randomly split into a training (80\% of data) 
and a testing (20\% of data) dataset, carefully avoiding  that data for the same subjects appeared  in both the training- and in the test-set.  
In order to accelerate the convergence of the models, we \textit{normalized}
each dimension of the feature vector \cite{Box:1989868}. 
Additionaly, we also used a leave-one-subject-out cross-validation strategy. Hence, 111 models for each personality trait were trained on 110-subject subsets, evaluating them against the remaining ones and finally averaging the results. The results obtained are not significantly different from the ones obtained using the random split 80\% vs 20\%. In the rest of the paper, we will discuss only the results obtained with the random split 80\% vs 20\%. 

\subsection{Feature Selection}

As an initial step, we carried out a \textit{Pearson correlation analysis} to
visualize and better understand the relations between variables in the feature
space. We found quite a large subset of features with strong mutual correlations
and another subset of uncorrelated features. Hence, there was room for feature
space reduction. We excluded using  \textit{principal component analysis} (PCA)
because the transformation it is based on produces new variables that are
difficult to interpret in terms of the original ones making  the interpretation
of the results more complex.

Therefore, we turned to a pipelined \textit{variable selection} approach, based
on \textit{feature ranking} and \textit{feature subset selection}, which was
perfomed using only data from the training set.
The metric used for feature ranking was the mean decrease in the \textit{Gini
coefficient of inequality}.
This choice was motivated because it outperformed other metrics, such as mutual
information, information gain and chi-square statistic with an average
improvement of approximately 28.5\%, 19\% and 9.2\% respectively
\cite{singh2010feature}.
The Gini coefficient ranges between $0$, expressing perfect equality in
predictive power and $1$, expressing maximal inequality in predictive power. The
feature with maximum mean decrease in Gini coefficient is expected to have the
maximum influence in minimizing the out-of-the-bag error. It is known in the
literature that minimizing the out-of-the-bag error results in maximizing common
performance metrics used to evaluate models (\emph{e.g.} accuracy, F1, AUC,
etc.) \cite{tuv2009feature}.

The feature selection procedure produced a reduced subset of 32 features from an
initial pool of about 500 features. Hence, we obtained a low-dimensional feature
space that makes our approach efficient to implement into mobile and multimedia
applications.

\subsection{Model Building}
We trained
a variety of classifiers: (i) an ensemble of tree classifiers based on a Random
Forest algorithm \cite{Breiman:2001:RF:570181.570182}, (ii) a Generalized Boosted
Model (GBM) \cite{Freund1997119}, (iii) Support Vector Machines with linear and
Gaussian radial basis kernels, and (iv) Neural Networks. The best solution of the
classification problem was found using an ensemble of tree classifiers based on
\emph{Random Forest} algorithm. In the rest of the paper, we report the performance results only for Random Forest.

Random forest
algorithm produces a combination of simple decision tree predictors, such that each tree is
dependent on the values of a random vector sampled independently with the same
distribution for all the classification trees in the
forest \cite{Breiman:2001:RF:570181.570182}. The decision boundary is formed
according to the margin function.
Given an ensemble of tree classifiers
$h_1(\mathbf{x}),h_2(\mathbf{x}),...,h_K(\mathbf{x})$ and if the training set is
drawn at random from the empirical distribution of the random vector $Y,
\mathbf{X}$ the margin function is defined as:

\begin{equation}
\begin{split}
mg(\mathbf{X},Y) = avg_k{I(h_k(\mathbf{X})=Y)} - \\
 \operatorname{max}_{j!=Y}avg_kI(h_k(\mathbf{X})=j),
\end{split}
\end{equation}
where $I(\cdot)$ is the characteristic function. The margin function measures
the distance between the average votes at $(\mathbf{X},Y)$ for the right class
and the average vote for any other class. For this model the generalization
error function is:
\begin{equation}
PE^* = P_{\mathbf{X},Y}(mg(\mathbf{X},Y)<0),
\end{equation}
where $P_{\mathbf{X},Y}$ is the probability over $\langle\mathbf{X},Y\rangle$
space.
For any event $A\subset\Omega$ of the feature space the characteristic function
$I(\cdot)$ of $A$ is:
\begin{equation}
I_A(x) = \begin{Bmatrix} 1 & {\iff (x\subset A)} \\ 0 & otherwise \end{Bmatrix} \begin{Bmatrix} 1 & {\iff \exists x} \\ 0 & otherwise \end{Bmatrix}
\end{equation}

Random Forests classifiers were trained with a stepwise increase of the number
of trees equal to the upper limit of $2^{11}$. Optimal number of trees for model
generalization as measured by mean misclassification rate for 10-fold cross-validation strategy is estimated to be 112 trees.

In order to find the final model, we trained a number of models and selected the
best one based on $\kappa$ metrics for the 10-fold validation strategy.
The Cohen's $\kappa$ measures pairwise agreement among a set of functions which
are making classification decisions with correction for an expected chance agreement
\cite{citeulike:2424556}:
\begin{equation}
\kappa=\frac{P(A)-P(E)}{1-P(E)}
\end{equation}
$\kappa=0$ if there is no agreement more than expected by chance following the
empirical distribution; while $\kappa=1$ when there is a $max$ agreement. $\kappa$ is
a state-of-the-art statistics about how significantly the classification model is different
from chance. Importantly, $\kappa$ is a more robust measure than the simple percent agreement, given that it 
takes into account chance agreement occurring without being too
conservative.

During the learning and model selection process we used a random sampling with
replacement to generate a new set of data for each fold from the basic training
set, and followed leave-one-out 10-fold cross validation scheme. We adopted
this strategy in order to prevent data overfitting and to deal with potential data
loss in cases where calls, sms and Bluetooth proximities existed in the real world but were not registered by the smartphone logger software. Our {\em structural risk minimization}, as opposed
to empirical risk minimization solution to prevent  data overfitting, was
incorporated by working with a regularization penalty into the learning process,
balancing the model's complexity against  training data fitting and by
sampling  the model training sets in such a  way that they mimic the empirical
distributions without most probable erroneous outliers.

Model parameter estimation selection was done iteratively on the
basis of our exploratory analysis, inferred knowledge of the relationships
between variables and model performance metrics ($\kappa$ and Accuracy). 
Confounding variables are identified but not removed from the dataset during training and
test phases. 

\section{Experimental Results}
The performance metrics used to evaluate our approach are: accuracy, $\kappa$, sensitivity,
and specificity. The recognition model based on random forest algorithm shows 90.68\%
accuracy on the training set and 72.28\% accuracy on the test set. 
In Table~\ref{tab:rfStatisticsComparison} we provide the final
stress recognition model performances on the test set 
along with their statistical significance \cite{citeulike:3849197}.

\begin{table}[ht]
\centering
\scalebox{1}{
\begin{tabular}{r|l}
  \hline
 Metric & Value \\ 
  \hline
               Accuracy & 0.7228 \\          
                95\% CI & (0.7051, 0.7399) \\
    No Information Rate & 0.6384 \\          
  P-Value [Acc $>$ NIR] & < 2.2e-16 \\       
 \\                                          
                  Kappa & 0.3752 \\          
 \\                                          
            Sensitivity & 0.5272 \\          
            Specificity & 0.8335 \\          
       'Positive' Class & ``stressed'' \\        
   \hline
\end{tabular}
}
\caption{Recognition Model Performance Metrics} 
\label{tab:rfStatisticsComparison}
\end{table}

Information about accuracy and $\kappa$  metrics distribution using 10-fold
cross validation strategy is provided in
Table~\ref{tab:rf10fCVStress}. As we can see, the distribution of the estimated performance metrics does not vary
substantially among  folds, signaling a good generalization despite  the
possible existence of heterogeneous data in each fold and the ``noise'' coming
from the resampling procedure.

\begin{table}[ht]
\centering
\scalebox{1}{
\begin{tabular}{r|l|l}
  \hline
& Accuracy & Kappa \\ 
  \hline
Min.   &0.6959   & 0.2995   \\ 
1st Qu.&0.7156   & 0.3535   \\ 
Median &0.7282   & 0.3817   \\ 
Mean   &0.7232   & 0.3684   \\ 
3rd Qu.&0.7312   & 0.3869   \\ 
Max.   &0.7404   & 0.4010   \\ 
   \hline
\end{tabular}
}
\caption{10-fold Cross-Validation Metrics} 
\label{tab:rf10fCVStress}
\end{table}

We also compared our approach based on combining multiple indicators with
simpler approaches  using as predictors (i) only personality traits, (ii) only
weather conditions, (iii) only activities inferred from mobile phone data, (iv)
a combination of personality traits and weather conditions, (v) a combination of
personality traits and activities inferred from mobile phone data, (vi) a
combination of weather conditions and activities inferred from mobile phone
data. Table~\ref{tab:subsets} reports accuracy, $\kappa$, sensitivity,
specificity and F1 for each approach.
In this table we also report the performance of (vii) a simple majority
classifier, which always returns the majority class as prediction (accuracy =
63.84\%).
Finally, we also ran experiments with three classes ("not stressed", "neutral",
"stressed"), with labels
 $-1$ for "not stressed", label $0$ for "neutral", and label $1$ for "stressed".
 The class "not stressed" included all the cases with scores $<4$, the class
 "neutral" included all the cases with scores $=4$, and the class "stressed"
 included all the scores $>4$. The sizes of the resulting three classes
are 42.83\% for "not stressed", 20.98\% for "neutral", and 36.15\% for
"stressed". The global accuracy obtained by our multifactorial model, 59.57\%,
significantly outperformed the performance of simple majority classifier, which
always returns the class "not stressed" as prediction.

\begin{table*}
\caption{Model Metrics Comparison for Feature Subsets}
\label{tab:subsets}
\centering
\scalebox{0.78}{
\begin{tabular}{l|rrrrrrrrrr}
  \hline
Model & Accuracy & Kappa & Sensitivity & Specificity & F1\\
  \hline
\bf{Our Multifactorial Model} & 72.28 & 37.52 & 52.72 & 83.35
 & 57.89 \\
  Baseline Majority Classifier & 63.84 & 0.00 & 100.00 & 0.00 & 0.00 \\
\hline
  Weather Only & 36.16 & 0.00 & 100.00 & 0.00 & 0.00 \\
  Personality Only & 36.16 & 0.00 & 100.00 & 0.00 & 0.00 \\
  Bluetooth+Call+Sms & 48.59 & 6.80 & 73.80 & 34.32 & 50.94 \\
  Personality+Weather & 43.55 & 2.96 & 81.90 & 21.83 & 51.20 \\
  Personality+Bluetooth+Call+Sms & 46.40 & 7.01 & 83.17 & 25.57
  & 52.88 \\
  Weather+Bluetooth+Call+Sms & 49.60 & -5.45 & 38.45 & 55.91 &
  35.55 \\
   \hline
\end{tabular}
}
\end{table*}

\section{Discussion}
The comparison among the performance of the various models in
Table~\ref{tab:subsets} provides convincing evidence that none of the features
sets (personality, weather, smartphone activity) considered alone is 
endowed with a good enough predictive power. This conclusion applies also to pairwise
combinations of the same features sets to the extent that neither
personality+smartphone activity, nor personality+weather, nor weather+smartphone
activity do any better than the majority classifier (accuracy=63.84\%). Interestingly, significant
improvements over the latter can only be obtained by the simultaneous usage of
the three features sets: our final model based on a Random Forest classifier
using 32-dimensional feature vectors obtained a 72.28\%
accuracy for our 2-class classification problem.

As pointed out in Section 2, some recent works have used mobile phones data for stress recognition \cite{bauer2012can, SanoPicard2013}. Bauer and Lukowicz \cite{bauer2012can} reported a 53\% of accuracy in detecting the transition from stressful periods (a two week exam session) to non-stressful periods (two weeks after the exam session). Our multifactorial approach outperforms the approach proposed by \cite{bauer2012can} altough a direct comparison may be not adequate given the different task. More recently, Sano and Picard reported an accuracy performance of 75\% using a combination of features from mobile phones and more obtrusive wearable sensors. However, the limited number of subjects (only 18) and the limited number of days (only 5) make the results preliminary.
Other approaches used video and audio features for stress recognition \cite{giakoumis2012real, Lu:2012:SDS:2370216.2370270}. For instance, StressSense, an application for stress detection from human voice, achieved a 76\% of accuracy in outdoor environments. However, this method depends on sound quality and it may pose privacy concerns for people perceiving voice recording and analysis as a threat to their privacy.
Hence, our performance of 72.28 shows that the proposed multifactorial approach is a reliable and less obtrusive alternative.

An investigation of the most important predictors of daily stress
reveales interesting associations. Table 1 reports the 32 features selected and used in our model ranked by their mean reduction in accuracy. 
All the personality traits contributes significantly in predicting the daily
stress variable. These results are interesting because the previous studies in
social psychology focused their analyses mainly on the associations between
stress and Neuroticism, Extraversion and Conscientiousness. Instead, our work
shows the important contribution played also by Agreeableness and Openness to Experience to
the automatic classification of daily stress. Moreover, these results open us
the possibility of creating a multi-step stochastic model in which  we first
estimate the personality and then we use those estimates as independent
variables for the daily stress recognition problem. Our current approach uses
self-reported information on personality and this strategy could be a limitation
for scaling to larger sample of users. However, recent studies showed that
personality traits may be reliable recognized from mobile phone data \cite{de2011towards, staiano2012friends, Chittaranjan:2013:MLS:2444516.2444522, Montjoye2013}.

With regard to weather, we found confirmation for the association between temperature
and stress. Moreover, significant effects of other meteorological variables --
humidity, visibility, and wind speed -- for
predicting daily stress were also found.

Regarding the mobile phone data,
it is interesting to note the  contribution of  proximity
features. Out of the selected 32 features, 11 features
are proximity ones, 6 comes from call data and 6 from sms data. In particular, an
interesting predictive role is played by the number of time intervals for which an id is seen.
The results obtained using proximity features seem to confirm previous findings in social psychology: in particular, the relevant role played by face-to-face interactions and by interactions with strong ties in determining the stress level of a subject \cite{Krackhardt1992}. For sure, this result requires further investigation.
In addition, two features capturing the entropy in proximity interactions are among the
selected ones. This finding seems to
confirm  results available in the  social psychological literature  about the associations between
stress and the richness/diversity of social interactions \cite{Goode1960}. 
Further confirmation to this conclusion comes from the similalry important role played by entropy-based call and sms features. 

The remaining selected features related to sms interactions are 
(i) the latency in replying to a text message, defined as the median 
time to answer a text message and (ii) the amount of sms communications (outgoing+incoming). 

\section{Implications and Limitations}
\label{sec:Implications}
Stress has become a major problem in our society. Ubiquitous connectivity, information overload,
increased mental workload and time pressure are all elements contributing to
increase general stress levels. While in some cases people may realize that they
are undergoing stressful situations, severe and chronic stress may be more difficult to
detect. Moreover, stress
may be considered the norm in a modern and demanding society. Nonetheless, while
slightly increased stress levels may be functional for productivity, prolonged
and severe stress can be at the source of several physical dysfunctions like
headache, sleep or immunological disorders, unhealthy behaviours such as smoking
and bad eating habits, as well as of psychological and relational problems.
Beside manifest social costs, stress also entails considerable financial costs
for our economies, which are estimated by the World Health Organization in
300 billion dollars a year for American enterprises, and 20 billion euro for
Europe ones, in terms of absenteeism and low productivity.

Our technology provides a cost-effective, unobtrusive, widely available and
reliable tool for stress recognition. It detects daily stress levels with a
72.28\% accuracy combining real life data from  different sources, such as
personality traits, social relationships (in terms of calls, sms and proximity interactions), and weather data. The development of a reliable stress
recognition system is a first but essential step toward wellbeing and
sustainable living, and its scope can be extended to different areas of
applicability. Providing people with a tool capable of gathering rich data about
real life, and transforming them into meaningful insights about stress levels,
paves the way to a new generation of context-aware technologies that can target
therapists, enterpises and common citizens.

This technology can inform the design of automatic
systems for the assessment and treatment of psychological stress. With such a
tool, therapists could monitor and record patients' daily stress levels, access
longitudinal data, identify recurrent or significant stressors and modulate
treatment accordingly.

In work environments, where stress has become a serious problem affecting productivity, leading to occupational issues and causing
health diseases, our system could be extended and employed for early detection
of stress-related conflicts and stress contagion, and for supporting balanced workloads.
Awareness is a first but crucial step to motivate people to change their
behaviour and take informed and concrete steps toward a healthy lifestyle and appropriate stress coping strategies. Mobile applications developed on the basis of our
technology could provide
feedback to increase people's awareness of their stress levels, alerts when they
reach a warning threshold, and suggest stress management and relaxation
techniques when appropriate.

However, our study has also some limitations. We can list the following ones: (i) our
sample comes from a population living in the same environment. Our subjects were
all married graduate students living in a campus facility of a major US
university; and (ii) the non-availability of proximity data concerning the
interaction with people not participating in the data collection, a fact that is
common to many other relevant studies and that has been also pointed out by
\cite{DBLP:conf/cscw/QuerciaECC12}.
The first problem is at least partially attenuated by the large variability of
the sample in terms of provenance and cultural background (in our sample we have subjects from 16 countries and from all the continents), which can be expected
to correspond to a wide palette of interaction behaviors that efficaciously
counterbalance the effects of living-place homogeneity.

\section{Conclusion}

The goal of this paper was to investigate the automatic recognition of people's
daily stress from three different sets of data: a) people activity, as detected
through their smartphones (data pertaining to transitory properties of
individuals); b) weather conditions (data pertaining to transitory properties of
the environment); and c) personality traits (data concerning permanent
dispositions of individuals). The problem was modeled as a 2-way classification
one. The results convincingly suggest that all the three types of data are
necessary for attaining a reasonable predictive power. As long as one of those
information sources is dropped, performances drop below those of the baselines.
Moreover, the distributional data for accuracy and $\kappa$  show the robustness
and generalization power of our multifactorial approach.

Taken together, and despite the limitations discussed above, our results not
only provide evidence that individual daily stress can be reliably predicted,
but they also point to the necessity of considering at the same time people's
transitory properties (smartphone activity), transitory properties of the
environment and information about stable individual characteristics.
For the sake of transitory individual properties, mobile phone usage patterns
have important advantage over alternative methods: they are less unobtrusive and
raise limited privacy problems as compared to, e.g., voice analysis or the
exploitation of data from physiological sensors. Moreover, and importantly,
automatic stress detection based on mobile phone data can take advantage of the
extensive usage and diffusion of these devices, it can be applied in several
real world situations and it can be exploited for a variety of applications that
are delivered by means of the same device. For example, applications used to
inform the design of clinical decision support systems or self-monitoring
applications of stress levels in work settings and in other daily life
situations, which allows people to identify personal stressors and enforces
their proactive role in stress prevention and management.

\section*{Acknowledgment}
This work was partially supported by Telecom Italia Semantics and Knowledge
Innovation Laboratory (SKIL) with research grant T.

\bibliographystyle{abbrv}
\bibliography{fp266-lepri}  
\end{document}